\documentclass[aps,pre,reprint,groupedaddress]{revtex4-2}
\usepackage{graphics}
\usepackage{physics}
\usepackage{color}

\begin{document}


\title{Bifurcation structure of the flame oscillation}


\author{Yuki Araya}
\affiliation{Department of Physics, Chiba University, Chiba 263-8522, Japan}
\author{Hiroaki Ito}
\affiliation{Department of Physics, Chiba University, Chiba 263-8522, Japan}
\author{Hiroyuki Kitahata}
\email[]{kitahata@chiba-u.jp}
\affiliation{Department of Physics, Chiba University, Chiba 263-8522, Japan}


\date{\today}

\begin{abstract}
A flame exhibits a limit-cycle oscillation, which is called ``flame flickering'' or ``puffing'', in a certain condition. We investigated the bifurcation structure of the flame oscillation in both simulation and experiment. We performed a two-dimensional hydrodynamic simulation by employing the flame sheet model. We reproduced the flame oscillation and investigated the parameter dependences of the amplitude and frequency on the fuel-inlet diameter. We also constructed an experimental system, in which we could finely vary the fuel-inlet diameter, and we investigated the diameter-dependences of the amplitude and frequency. In our simulation, we observed the hysteresis and bistability of the stationary and oscillatory states. In our experiments, we observed the switching between the stationary and oscillatory states. As fluctuations can induce the switching in the bistable system, switching observed in our experiments suggested the bistability of the two states. Therefore, we concluded that the oscillatory state appeared from the stationary state through the subcritical Andronov-Hopf bifurcation in both the simulation and experiment. The amplitude was increased and the frequency was decreased as the fuel-inlet diameter was increased. In addition, we visualized the vortex structure in our simulation and discussed the effect of the vortex on the flame dynamics.
\end{abstract}


\maketitle


\section{introduction}
Periodic behaviors can be observed in biological\cite{Murray2002,pikovsky2001}, chemical\cite{Mikhailov2017,kuramoto2003}, and hydrodynamical\cite{Ball2011,Lappa2009} systems. They have been understood as nonlinear oscillations and have been studied in relation to the bifurcation, synchronization, and pattern formation. One of the simple phenomena that are understood as limit-cycle oscillations is the flame oscillation, which is often called ``flame flickering'' or ``puffing''\cite{Chamberlin1928,Faraday1861}. One of the authors reported that a single candle burns in a stationary manner, whereas the merged flame on a bundle of three-or-more candles oscillates at a typical frequency of around 10 Hz\cite{Kitahata2009}. The study reported that the flame oscillation emerges with a greater amount of fuel supply than a threshold. Flame oscillations are generally observed in two types of diffusion flames: a pool flame, where the fuel vapor is introduced into the system through the evaporation from a liquid pool\cite{Cetegen1993}, and a jet flame, where the fuel stream is supplied via a gaseous jet\cite{Hamins1992}. Note that candle flames are classified as a pool flame. In both systems, the frequency in the oscillatory state is universally proportional to $d^{-1/2}$ over a wide range of the fuel-inlet diameter $d$\cite{Xia2018}. The flame oscillation has attracted the attention of many researchers not only of fundamental sciences but also in the industrial fields, since the oscillatory combustion leads to more emission of pollutants, such as unburned hydrocarbons and soot, than stationary combustion.

The flame oscillation has been intensively studied based on hydrodynamics\cite{Hamins1992,Xia2018,Fujisawa2018,Carpio2012,Durox1997,Zhou2001,Yang2019,Bunkwang2020,Chen2019,Moreno-Boza2016,Tokami2021}. Xia et al. reproduced $d$-dependence of the frequency, proportional to $d^{-1/2}$ irrespective of fuel types, with a theoretical analysis based on vortex dynamics. Fujisawa et al. experimentally measured the velocity field around the oscillating flame with particle image velocimetry, performed ``invalid velocity vector analysis'' based on the proper orthogonal decomposition, and observed a strong correlation between vortex structures and flame shapes\cite{Fujisawa2018}.

From the viewpoint of dynamical systems, the emergence of the flame oscillation was considered to be a transition from a stationary state to an oscillatory state\cite{Strogatz1994}. Many other types of transitions with regard to flame dynamics have also been investigated: transition between the flame oscillations with and without ``pinch-off''\cite{Carpio2012}, transition between periodic, quasiperiodic, and chaotic oscillations with a rotating cylindrical burner\cite{Gotoda2009}, and switching between the axisymmetric and asymmetric shapes of oscillating flames\cite{Cetegen2000,Bunkwang2020}. It is also reported that two-or-more oscillating flames can synchronize according to the spatial arrangement of fuel inlets\cite{Kitahata2009}. The transitions between various synchronization modes have been observed\cite{Kitahata2009,Bunkwang2020,Gergely2020,Manoj2018,Okamoto2016,Chen2019,Yang2019,Dange2019,Changchun2019,Fujisawa2020,Okamoto2016,Forrester2015,Manoj2019,Manoj2021} and have also been investigated using mathematical models\cite{Kitahata2009,Gergely2020}.

When we consider transitions in terms of bifurcation phenomena, investigation of bifurcation structures allows us to evaluate the validity of mathematical models and can provide insights into the mechanism of the phenomena. However, most studies of flame oscillations have focused on reproducing the phenomena, and only a few studies have focused on the bifurcation structures. Moreno-Boza et al. theoretically investigated the chemo-hydrodynamic instability of the stationary state\cite{Moreno-Boza2016} with the flame sheet model, in which the reaction rate of the combustion is sufficiently high and the Lewis number, the ratio of the thermal diffusion coefficient to the mass diffusion coefficient, is unity\cite{Mahalingam1999}. They compared the result of linear stability analysis with that from hydrodynamic simulation under the assumption that the flame oscillation occurs through a supercritical Andronov-Hopf bifurcation from the stationary state, and reported some discrepancies, which were left for future investigation\cite{Moreno-Boza2016}. The Andronov-Hopf bifurcation is one of the well-known bifurcation structures where the oscillatory state with a finite frequency appears from the stationary state. This bifurcation is classified into two types: the supercritical and subcritical Andronov-Hopf bifurcations\cite{Strogatz1994,Wiggins2003}. Considering that only the finite frequency has been reported for the flame oscillation\cite{Cetegen1993,Hamins1992,Xia2018,Chen2019}, the bifurcation structure of the flame oscillation should be classified into the supercritical or subcritical Andronov-Hopf bifurcation. To further understand the flame oscillation, it is essential to identify the bifurcation structure embedded in this phenomenon.

Thus, in the present study, we investigated the bifurcation structure of the flame oscillation in both simulation and experiment. We performed a two-dimensional hydrodynamic simulation employing the flame sheet model, and investigated the dependences of the amplitude and frequency on the fuel-inlet diameter. We considered that our two-dimensional simulation model includes the essential factors, such as the combustion reaction, buoyancy, and convective heat transfer, to reproduce the flame behavior. In order to confirm the simulation result, we also constructed an experimental system in which we could finely vary the fuel-inlet diameter, and investigated the dependences of the amplitude and frequency on the diameter. We then identified the bifurcation structure from the results of the simulation and experiment. We also visualized the vortex structure in our simulation and discussed the effect of the vortex on the flame dynamics based on the phase description of the flame oscillation.

\section{simulation setup}
To reproduce the flame oscillation, we considered the hydrodynamics and the combustion dynamics for the gas, which is composed of ethanol, oxygen, and other non-reactive compounds. We performed two-dimensional hydrodynamic simulation assuming that the gas around the flame is a compressible Newtonian fluid. We used the equation of continuity and the Navier-Stokes equation,
\begin{align}
  \partial_t\rho+\partial_\alpha\left(\rho v_\alpha\right)=0,\label{quad:1}
\end{align}
\begin{align}
  \partial_t(\rho v_\alpha)+\partial_\beta\left(\rho v_\beta v_\alpha\right)=&\frac{1}{Re}\left(\partial_\beta\partial_\beta v_\alpha+\frac{1}{3}\partial_\alpha\partial_\beta v_\beta\right)\nonumber\\
  &-\partial_\alpha P-\frac{1}{Fr}(\rho-\rho_0)\delta_{\alpha 2},\label{quad:2}
\end{align}
which are nondimensionalized by the characteristic velocity, length, and density scales. Here, $t$ is the time, $\partial_t$ is the partial differential operator with respect to $t$, $x_\alpha\,(\alpha=1,2$) is the spatial coordinate, $\partial_\alpha$ is the partial differential operator with respect to $x_\alpha$, and $\delta_{\alpha\beta}$ is the Kronecker delta. Here and henceforth, we use the subscripts $\alpha$ and $\beta$ to denote the index of the spatial coordinate and use the summation convention. $\rho$, $v_\alpha$, and $P$ are the density, velocity, and pressure field, respectively. $Re$ is the Reynolds number, the ratio of the inertial force to the viscous force, and $Fr$ is the Froude number, the ratio of the inertial force to the gravitational force. $\rho_0$ is the density of the gas with no combustion.

In order to consider the ethanol combustion simply, we adopted the flame sheet model\cite{Mahalingam1999}. In this model, the temperature and the mass fractions of the ethanol, oxygen, and other compounds commonly depend on a single variable, ``mixture fraction'' $Z$\cite{Williams1985}, and it follows the advection-diffusion equation,
\begin{align}
  \partial_tZ+v_\alpha\partial_\alpha Z=\frac{1}{RePr}\frac{1}{\rho}\partial_\alpha\partial_\alpha Z.\label{quad:3}
\end{align}
Here, $Pr$ is the Prandtl number, the ratio of the kinematic viscosity to the thermal diffusivity. The temperature field $T$ depends on $Z$, as
\begin{align}
  T=\left\{
    \begin{array}{ll}
      T_0+(T_1-T_0)Z/Z_{\rm st},&Z<Z_{\rm st},\\
      T_0+(T_1-T_0)(1-Z)/(1-Z_{\rm st}),&Z\geq Z_{\rm st},
    \end{array}
  \right.\label{quad:5}
\end{align}
where $Z_{\rm st}$ is the stoichiometric mixture fraction, $T_0$ is the temperature with no combustion, and $T_1$ is the highest temperature in the system. $T_1$ and $Z_{\rm st}$ should depend on the actual fuel type and the initial and/or boundary condition of the mass fraction. In order to ignore the sound wave, we adopted the low Mach number approximation, assuming that the Mach number, which is the ratio of the flow velocity to the sound velocity, in the flame oscillation was negligibly smaller than unity\cite{Rehm1978}. We also assumed that the gas around the flame is the ideal gas, where the dependence of the density on the mass fraction was ignored. Under these approximations, $\rho$ and $T$ satisfy the equation of state for the ideal gas,
\begin{align}
  \rho T=\rho_0T_0.\label{quad:4}
\end{align}
\begin{figure*}[tb]
  \includegraphics{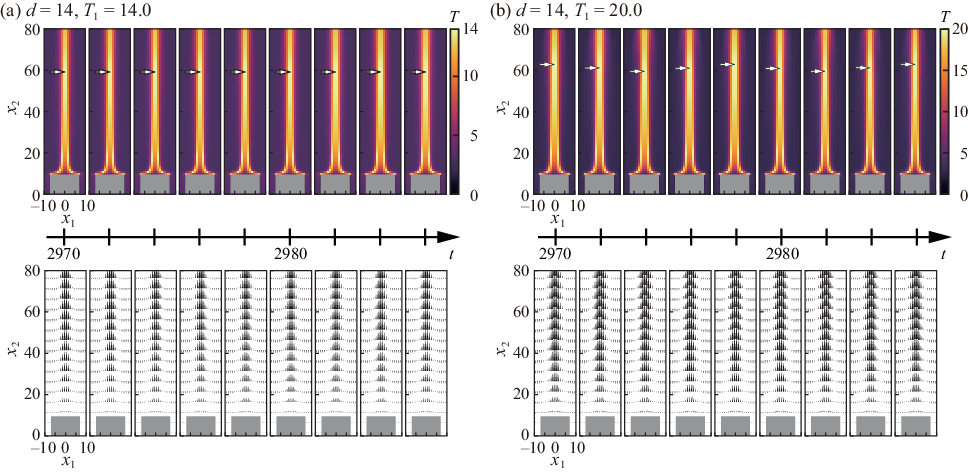}
  \caption{
    Sequential snapshots of the temperature field $T$ (upper panels) and the velocity field $v_\alpha$ (lower panels) after a sufficiently long time obtained by the numerical calculation for (a) $d=14,T_1=14$ and (b) $d=14,T_1=20$. The wick region is represented by gray. The top position of the flame is indicated by the arrow in each snapshot for the upper panels.
  \label{fig:6}}
\end{figure*}

For numerical simulation, we considered a rectangular region, $-X_1\leq x_1\leq X_1$, $0\leq x_2<X_2$, and defined the floor as $x_2=0$. We also defined the region for the wick as $-d/2\leq x_1\leq d/2$, $0\leq x_2\leq h$, where $d$ and $h$ are the width and height of the wick, respectively. As we assumed a bilateral symmetry of the system, we set the symmetry axis to $x_1=0$. The actual calculation region, half of the system, was thus $0\leq x_1\leq X_1$, $0\leq x_2\leq X_2$. The nonslip boundary condition for the velocity, $v_\perp=v_\parallel=0$, was adopted for the floor and surfaces of the wick. Here, $v_\perp$ and $v_\parallel$ are the velocity components in the directions perpendicular and parallel to the boundary, respectively. A slip boundary condition for the velocity, $v_\perp=\partial_\perp v_\parallel=0$, was adopted for the symmetry axis. Here, $\partial_\perp$ and $\partial_\parallel$ are the partial differential operators with respect to $x_\perp$ and $x_\parallel$, which are the spatial coodinates in the directions perpendicular and parallel to the boundary, respectively. The velocity at $x_1=X_1$ and $x_2=X_2$ follows the Neumann boundary condition, $\partial_\perp v_\perp=\partial_\perp v_\parallel=0$. The pressure at the symmetry axis, the floor, and the surfaces of the wick was determined so that it satisfies the Navier-Stokes equation in Eq.~(\ref{quad:2}). The pressure at $x_1=X_1$, and $x_2=X_2$ follows the Dirichlet boundary condition, $P=0$. The Neumann boundary condition for the mixture fraction, $\partial_\perp Z=0$, was adopted for the symmetry axis, the floor, and the side of the wick. Considering that the fuel evaporates from the upper surface of the wick, the Dirichlet boundary condition for the mixture fraction, $Z=1$, was adopted for the upper surface of the wick. The mixture fraction at $x_1=X_1$ and $x_2=X_2$ follows the Dirichlet boundary condition, $Z=0$. To solve the equation of continuity and the Navier-Stokes equation, we used the fractional step method\cite{J.Kim2013} and calculated the variables on a staggered grid. We used an explicit method for the advection-diffusion equation to calculate the time evolution of $Z$. The convective terms were handled by the upwind scheme and the other terms were handled by the central difference scheme. We set the time step $dt=0.01$ and the spatial mesh $dx_1=dx_2=1$. The size of the calculation region was set as $X_1=50$ and $X_2=100$. $h$ was set constant at 10 and $d$ was varied, which could only be an even number due to the bilateral symmetry in a discrete grid. Dimensionless parameters were set as $Re=\varrho UL/\mu=21.58$, $Pr=\mu c/\kappa=0.7099$, and $Fr=U^2/gL=2.041$, where $L=0.002\,{\rm m}$, $U=0.2\,{\rm m/s}$, and $\varrho=1.0\,{\rm kg/m^3}$ were the characteristic length, velocity, and density, respectively. $\mu=1.853\times 10^{-5}\,{\rm Pa\cdot s}$, $\kappa=0.026\,{\rm W/(m\cdot K)}$, and $c=1000\,{\rm J/(kg\cdot K)}$ were the viscosity, thermal conductivity, and specific heat capacity of the air, respectively, and $g=9.8\,{\rm m/s^2}$ was the gravitational acceleration. We set $Pr=0.1789$ for the upper surface of the wick so that the diffusion may be great enough to provide sufficient amount of ethanol from the upper surface of the wick. The other simulation parameters were set as follows: $\rho_0=1.18$, $T_0=3$, and $Z_{\rm st}=0.2199$. $T_1$ was varied finely as a control parameter. We performed the simulations for $d=8,\,10,\,12$, and $14$ in the following procedure. First, we calculated 1,000 steps for $T_1=T_0=3$ with $dt=0.1$, which corresponded to the diffusion process of ethanol into the air before ignition. The initial conditions for $v_\alpha$ and $Z$ were $v_1=v_2=Z=0$ in the whole region. In this situation, buoyancy should not work and convective flow should not occur since we assumed that the density depends only on the temperature as in Eq.~(\ref{quad:4}). Next, we reset $T_1=12$, which corresponded to the ignition, and set $t=0$ at this instant. We calculated till $t=3,000$ with $dt=0.01$, and the variables in the final state of this calculation were set as an initial condition of the following calculation with different $T_1$. We increased $T_1$ from 12 to 24 in increments of 0.5, and then decreased from 24 to 12 in decrements of 0.5. We reset $t=0$ when we varied $T_1$, and calculated till $t=3,000$ with $dt=0.01$ for each $T_1$.

\section{Simulation results}
The sequential snapshots of the temperature field $T$ and the velocity field $v_\alpha$ after a sufficiently long time are shown in Fig.~\ref{fig:6}. The stationary state, in which the flame height was constant, is shown in Fig.~\ref{fig:6}(a) for $d=14$ and $T_1=14$, while the oscillatory state, in which the flame height oscillated, is shown in Fig.~\ref{fig:6}(b) for $d=14$ and $T_1=20$. In the flame sheet model, we can uniquely define the contour of the flame as $Z=Z_{\rm st}$, where the chemical reaction occurs. Therefore, the flame height is defined as the $x_2$-coordinate of the top position of the flame, which is indicated by an arrow in each snapshot in Fig.~\ref{fig:6}. As shown in the velocity fields, we observed a strong upward flow above the wick in both states.

\begin{figure}[tb]
  \includegraphics{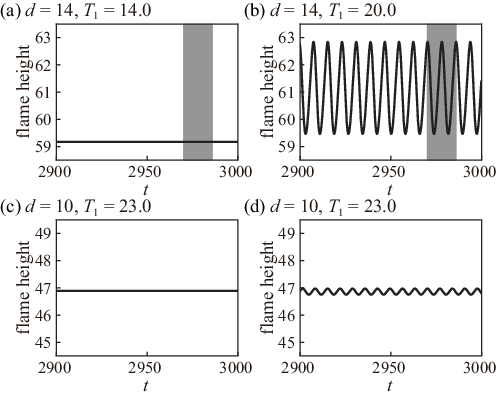}
  \caption{
    Time series of the flame height, which is defined as the top position indicated by the arrows in Fig.~\ref{fig:6}, for (a) $d=14,T_1=14$, (b) $d=14,T_1=20$, and (c,d) $d=10,T_1=23$. Initial conditions were different for (c) and (d). The time ranges indicated by the shaded regions in (a) and (b) correspond to those for the snapshots in Fig.~\ref{fig:6}(a) and (b), respectively.
  \label{fig:7}}
\end{figure}

The time series of the flame height for the parameters in Fig.~\ref{fig:6}(a) and (b) are shown in Fig.~\ref{fig:7}(a) and (b), respectively. For $d=10$ and $T_1=23$, either the stationary state shown in Fig.~\ref{fig:7}(c) or the oscillatory state shown in Fig.~\ref{fig:7}(d) was observed depending on the initial condition. The stationary and oscillatory states were observed when $T_1$ was increased and decreased, respectively, which indicates the bistability of the two states.

\begin{figure}[tb]
  \includegraphics{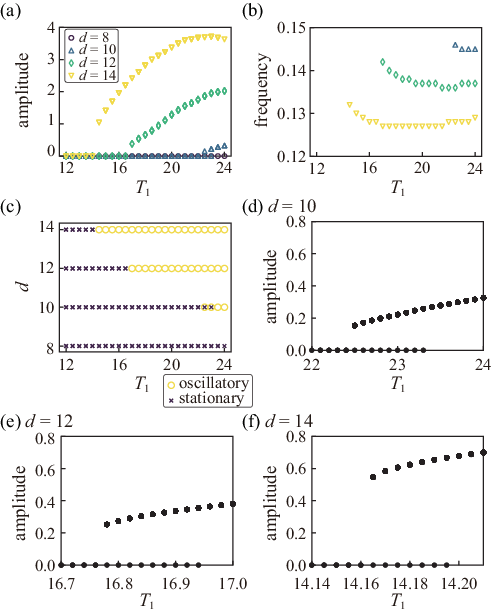}
  \caption{
    (a) Amplitude and (b) frequency of the oscillation in the flame height depending on $T_1$ for various $d$. (c) Phase diagram to distinguish the stationary and oscillatory states on the $d$-$T_1$ plane. Amplitudes of the oscillations in the flame heights for $d=$ (d) $10$, (e) $12$, and (f) $14$, when $T_1$ was varied finely.
  \label{fig:8}}
\end{figure}

In the oscillatory state, the difference between the maximum and minimum values of the flame height was employed for the amplitude, and the frequency at which the Fourier spectrum took the maximum within a range of 0.05--0.20 was employed for the frequency of the oscillation. In the stationary state, the amplitude should be 0. $T_1$-dependences of the amplitude and frequency for each $d$ are shown in Fig.~\ref{fig:8}(a) and (b), respectively. The phase diagram on the $d$-$T_1$ plane is shown in Fig.~\ref{fig:8}(c). The stationary state, in which the amplitude was zero, was observed for small $T_1$ and $d$. In contrast, the oscillatory state, in which the amplitude was finite, was observed for large $T_1$ and $d$. In the oscillatory state, the amplitude was increased and the frequency was decreased as $d$ was increased for fixed $T_1$. In order to investigate the bifurcation structure, we varied $T_1$ finely around the bifurcation point for each $d$. For example, since the bifurcation should occur in $16.5\leq T_1\leq17$ for $d=12$ as shown in Fig.~\ref{fig:8}(c), we increased $T_1$ from 16.5 and decreased $T_1$ from 17 in increments and decrements of 0.02, respectively. For the other $d$, $T_1$ was varied in the same way (between 22 and 24 in increments/decrements of 0.1 for $d=10$, and between 14 and 14.5 in increments/decrements of 0.005 for $d=14$). $T_1$-dependence of the amplitude is shown in Fig.~\ref{fig:8}(d), (e), and (f) for $d=10$, 12, and 14, respectively. The system was in the stationary or oscillatory state depending on how the parameter had been varied before reaching the same $T_1$, which clearly indicated the existence of hysteresis and bistability. These results suggested that the oscillatory state appeared from the stationary state through the subcritical Andronov-Hopf bifurcation.

\section{Experimental setup}
\begin{figure}[tb]
  \includegraphics{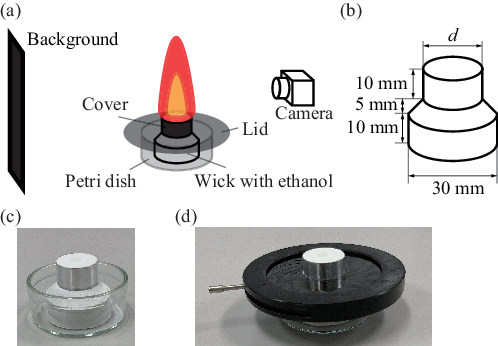}
  \caption{
    (a) Experimental setup. (b) Design of the wick made of diatomite. (c) Wick, whose side surface was covered with a tape, on a petri dish. (d) The wick with a lid.
  \label{fig:1}}
\end{figure}

To compare with the parameter dependences observed in the simulation, we performed experimental observations of the flame for various wick diameters. Figure~\ref{fig:1}(a) shows the experimental setup. We prepared wicks by cutting diatomite (B161WH, Soil, Japan) into the shape depicted in Fig.~\ref{fig:1}(b) with a milling machine (monoFab SRM-20, Roland, Japan). The wick diameter $d$ was varied from 8 mm to 28 mm in increments of 2 mm as a control parameter. In order to prevent evaporation of the fuel from the side surface of the wick, we covered the side surface with an aluminum tape (width: 10 mm, thickness: 0.1 mm, B07SF5GCHT, TeenitorJP, China), placed it in a glass Petri dish (diameter: 45.5 mm, height: 18.8 mm) (Fig.~\ref{fig:1}(c)), and covered it with a lid (D50S, Thorlabs, USA)  (Fig.~\ref{fig:1}(d)). We performed the experiments in the following procedure. First, we immersed the wick covered with the tape in ethanol (Wako Pure Chemicals, Japan), and degassed it for $10\,{\rm min}$ in a vacuum chamber in order to let it absorb the ethanol. Next, we prepared the setup shown in Fig.~\ref{fig:1}(a), and ignited the wick in a darkroom. At the same time, we started recording videos of the flame for $90\,\rm s$ using a high-speed camera (300.16 fps, $540\times256$ pixels, STC-MBS43U3V, Omron Sentech, Japan) equipped with an objective lens (L-600-12, Hozan, Japan). We performed the experiments six times for each $d$ with the above procedure. The wicks were reused up to two times.

\section{Experimental results}
\begin{figure}[tb]
  \includegraphics{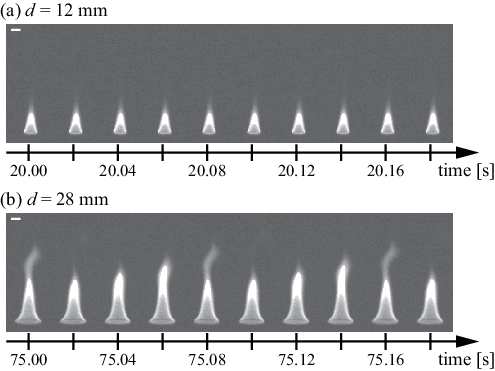}
  \caption{
    Sequential snapshots of the flame taken every $0.02\,\rm s$ obtained by experiments for $d=$ (a) $12\,{\rm mm}$ and (b) $28\,{\rm mm}$. Scale bar: $10\,{\rm mm}$.
  \label{fig:2}}
\end{figure}

The sequential snapshots of the flame taken every $0.02\,\rm s$ are shown in Fig.~\ref{fig:2}. The stationary state, in which the flame height was constant (Fig.~\ref{fig:2}(a)), and the oscillatory state, in which the flame height oscillated (Fig.~\ref{fig:2}(b)), were observed for $d=12\,\rm mm$ and $28\,\rm mm$, respectively.

\begin{figure*}[tb]
  \includegraphics{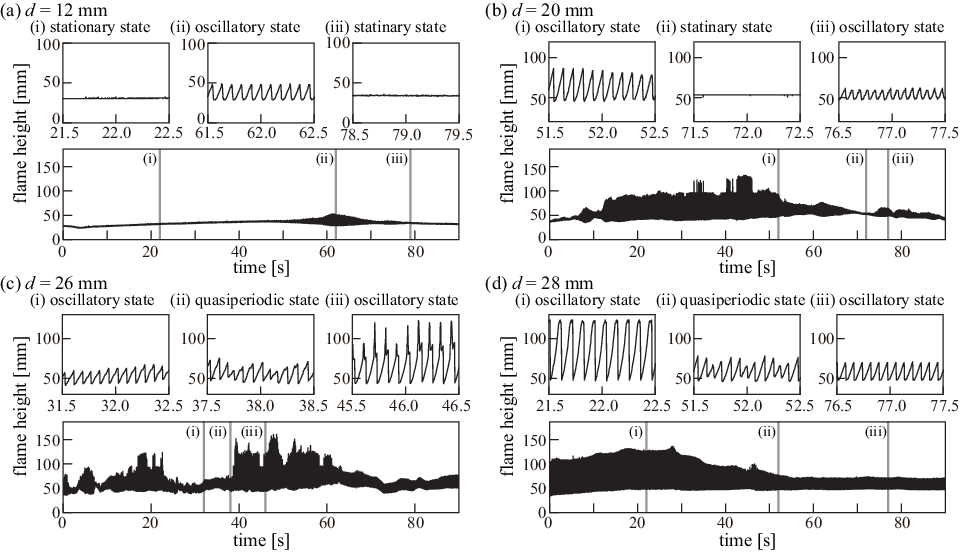}
  \caption{
    Time series of the flame height for $d=$ (a) $12\,{\rm mm}$, (b) $20\,{\rm mm}$, (c) $26\,{\rm mm}$, and (d) $28\,{\rm mm}$. The upper panels (i)--(iii) for each $d$ are the detailed plots for the thin shaded regions (i)--(iii) in the lower panel.
  \label{fig:3}}
\end{figure*}

In order to quantify the dynamics of the flame shape, we performed the binarization on the brightness of each pixel in the recorded videos, and defined the vertical component of the top position of the flame as the flame height. The time series of the flame height are shown in Fig.~\ref{fig:3}. We observed the stationary state, the oscillatory state, and the quasiperiodic state. Here, the quasiperiodic oscillation was composed of two-or-more incommensurate frequencies. We also observed the switching among these states.

\begin{figure}[tb]
  \includegraphics{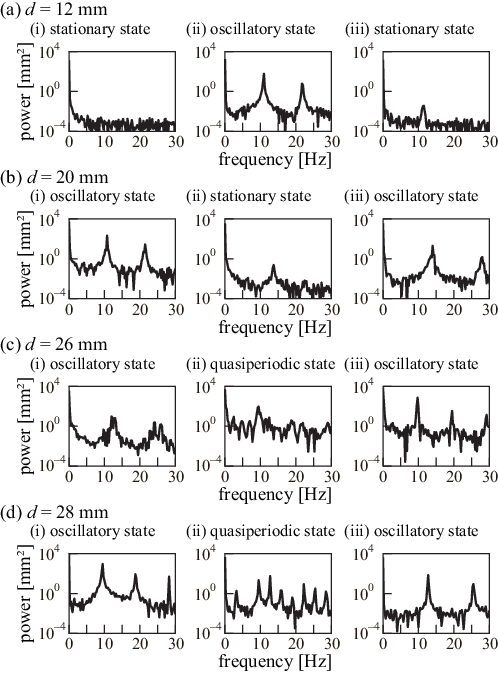}
  \caption{
    Power spectra of the flame height for $d=$ (a) $12\,{\rm mm}$, (b) $20\,{\rm mm}$, (c) $26\,{\rm mm}$, and (d) $28\,{\rm mm}$. The spectra of (i)--(iii) correspond to the time series in Fig.~\ref{fig:3}.
  \label{fig:12}}
\end{figure}

In order to determine the frequency of the flame oscillation, we performed the Fourier transform on the time series of the flame height. The power spectrum corresponding to each time series shown in Fig.~\ref{fig:3} is shown in Fig.~\ref{fig:12}. In the oscillatory state, there was a peak around $\rm 10\,Hz$ and its multiples. In the stationary state, there was no peak (Fig.~\ref{fig:12}(a)(i)) or a much smaller peak than that in the oscillatory state (Fig.~\ref{fig:12}(a)(iii) and (b)(ii)). In the quasiperiodic state, there were some peaks at equal intervals around the highest peak (Fig.~\ref{fig:12}(c)(ii) and (d)(ii)).

Considering that the state of the system could vary over time, we measured the amplitude and frequency of the flame oscillation every $1\,\rm s$ in the time course. We performed the Fourier transform on the time series of the flame height during the $4\,\rm s$ window around the considered time, and the frequency at which the Fourier spectrum took the maximum in the frequency range between $5\,\rm Hz$ and $30\,\rm Hz$ was employed for the frequency of the oscillation of the flame height. The amplitude of the flame oscillation was defined as the difference between the local maximum and minimum values in the time course of the flame height. Since the frequency of the flame oscillation was about $10\,\rm Hz$, we detected the local maximum values in the time course of the flame height so that the time interval between them was greater than 0.05~s. The minimum values between the local maximum values were employed for the local minimum values. We averaged the amplitude over the time window of $1\,\rm s$ at every $1\,\rm s$ in its time course. Note that the quasiperiodic state is classified into the oscillatory state through this procedure.

\begin{figure*}[tb]
  \includegraphics{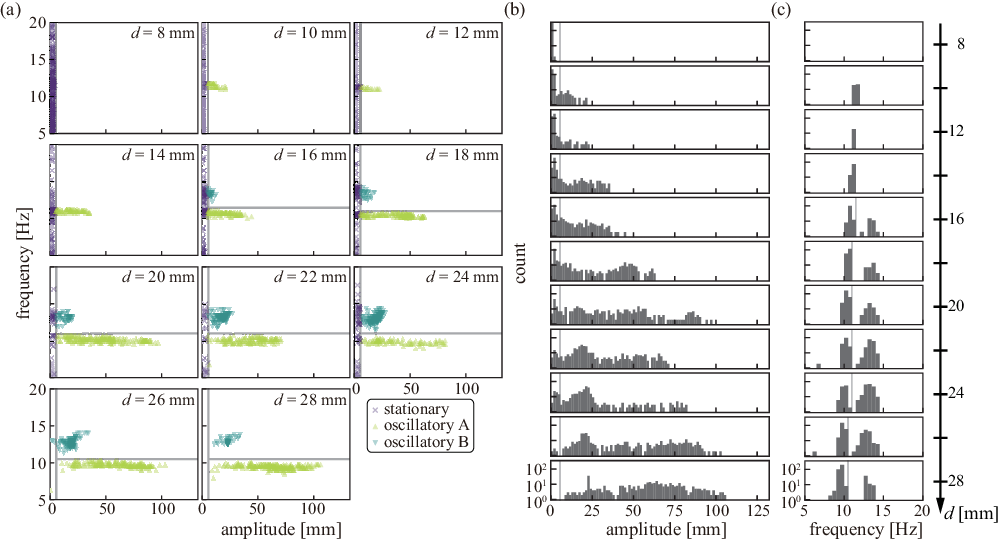}
  \caption{
    (a) Scatter plots of the amplitude and frequency of the flame height for various $d$. Threshold values of the amplitude and frequency are indicated by gray vertical and horizontal lines, respectively. Histograms of (b) the amplitude and (c) the frequency for each $d$ with a logarithmic vertical axis. The oscillatory and stationary states were distinguished by a threshold value of the amplitude indicated by a gray vertical line (${\rm amplitude}=5.5\,{\rm mm}$) in (a) and (b). The histograms of the frequency were prepared using the data only for the oscillatory state (${\rm amplitude}\geq5.5\,{\rm mm}$). Oscillatory states A and B were distinguished by a threshold value of the frequency indicated by a gray horizontal line in (a) and a gray vertical line in (c). The classification by the thresholds is indicated by different colors and marks in (a).
  \label{fig:13}}
\end{figure*}

We detected 60 pairs of the amplitude and frequency for 20--79~s from the ignition time in each experiment. The scatter plots of the amplitude and frequency for each $d$ are shown in Fig.~\ref{fig:13}(a). In the stationary state, the amplitude was about $0.0\,{\rm mm}$ and the frequency was distributed over a wide range, and in the oscillatory state, the amplitude was finite and the frequency was almost constant at one or two values. Since both the stationary and oscillatory states were observed for the same $d$, they may be bistable. Since the frequency typically took one of the two values in the oscillatory state for $16\leq d\leq28\,{\rm mm}$, two-types of oscillatory states exist: oscillatory state A, where the frequency is lower (Fig.~\ref{fig:12}(b)(i), (c)(iii), and (d)(i)), and oscillatory state B, where the frequency is higher (Fig.~\ref{fig:12}(b)(iii), (c)(i), and (d)(iii)). We also observed the switching between oscillatory states A and B through the stationary state or quasiperiodic state as shown in Figs.~\ref{fig:3} and \ref{fig:12}.

In order to distinguish the stationary and oscillatory states, we defined a common threshold value of the amplitude as $5.5\,{\rm mm}$ for all $d$ as shown in Fig.~\ref{fig:13}(a) and (b). The histograms of the frequency in the oscillatory state for each $d$ are shown in Fig.~\ref{fig:13}(c). In order to distinguish oscillatory states A and B, we defined a threshold value of the frequency where the count was minimum between the two peaks for each $d$ as shown in Fig.~\ref{fig:13}(a) and (c). In case there were multiple candidates where the counts were minimum between the two peaks, we employed the value closest to the peak of oscillatory state A. The classification by the thresholds is shown with different markers in Fig.~\ref{fig:13}(a), and the classification was appropriate.

\begin{figure}[tb]
  \includegraphics{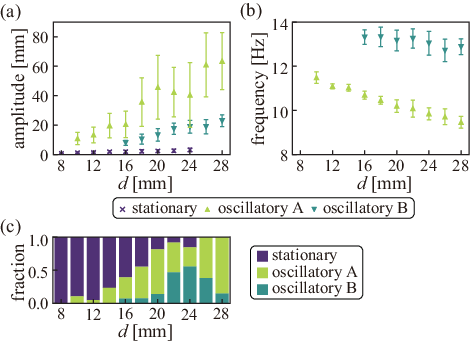}
  \caption{
    Mean values of (a) the amplitude and (b) the frequency in each state, i.e., stationary state, oscillatory state A, and oscillatory state B, for various $d$. Error bars indicate the standard deviation. (c) Fraction of the time duration for each state at different $d$.
  \label{fig:5}}
\end{figure}

The results of the mean and standard deviation of the amplitude and frequency for each state are summarized in Fig.~\ref{fig:5}(a) and (b). As $d$ was increased, the amplitudes increased in both oscillatory states A and B, and the frequency decreased in oscillatory state A. In contrast, the frequency did not clearly decrease in oscillatory state B. The fraction of time duration for which each state was observed is shown in Fig.~\ref{fig:5}(c). As $d$ was increased, the fraction of the stationary state decreased and that of the oscillatory state increased. The fraction of oscillatory state B was maximum at $d=24\,{\rm mm}$.

\section{discussion}
\begin{figure}[tb]
  \centering
  \includegraphics{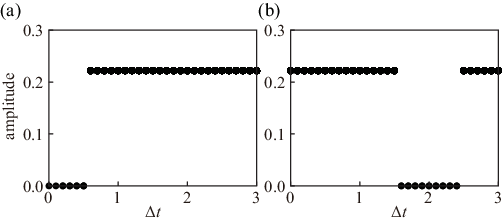}
  \caption{
    Amplitude of the oscillation in the flame height depending on $\Delta t$ for $d=10,T_1=23$. The Prandtl number $Pr$ at the upper surface of the wick from $Pr=0.1789$ into $Pr=0.2236$ at $t=3000$ and changed back into $Pr=0.1789$ at $t=3000+\Delta t$ under the conditions to (a) Fig.~\ref{fig:7}(c) and (b) Fig.~\ref{fig:7}(d).
  \label{fig:15}}
\end{figure}

Our experimental results shown in Fig.~8 suggested that the stationary and oscillatory states were bistable, and the noise, such as gentle convective flows and fluctuation of fuel supply, may induce the switching between these states. Therefore, We checked whether a fluctuation induced the switching in the simulation. As the fluctuation of fuel supply, we changed the Prandtl number $Pr$ at the upper surface of the wick from $Pr=0.1789$ into $Pr=0.2236$ at $t=3000$ and changed back into $Pr=0.1789$ at $t=3000+\Delta t$. We show $\Delta t$-dependence of the amplitude for $d=10$, $T_1=23.0$ after a sufficiently long time in Fig.~\ref{fig:15}. When the time duration $\Delta t$ of the perturbation was greater than $\Delta t=0.5$, the system switched from the stationary state to the oscillatory state (Fig.~\ref{fig:15}(a)). The system switched from the oscillatory state to the stationary state when $1.6\leq\Delta t\leq2.4$ (Fig.~\ref{fig:15}(b)). Therefore the switching between the stationary and oscillatory states could be reproduced in the simulation.
  
As $d$ was increased, the states observed in our experiments typically changed in the following order: stationary state, bistable state, and oscillatory state. Thus, the oscillatory state may appear from the stationary state through the subcritical Andronov-Hopf bifurcation, which was in agreement with our simulation results. In the simulation (Fig.~\ref{fig:8}), the bistable region and the oscillatory amplitude therein were so small that the parameter dependence of the amplitude may be similar to the one under the assumption of the supercritical Andronov-Hopf bifurcation. Therefore, we suggest that the hysteresis and bistability should be adequately evaluated to determine the bifurcation points.

$d$-dependences of the amplitude and frequency obtained in the simulation were qualitatively consistent with those for oscillatory state A obtained in the experiments. In a previous study, Xia et al. reproduced the power law, in which the frequency was proportional to $\sqrt{g/d}$, with the theoretical analysis based on vortex dynamics\cite{Xia2018}. Following their results, the frequency for $d\approx10^1\,{\rm mm}$ and $g\approx10^1\,{\rm m/s^2}$, typical parameters in our experiments, was almost $10^1\,{\rm Hz}$, which was at an order consistent with that of our experimental results. Although this power law was observed over three orders of magnitude of $\sqrt{g/d}$, the range of $\sqrt{g/d}$ in our experiments, which was limited near the bifurcation point as $18.7\,{\rm s^{-1}}\leq\sqrt{g/d}\leq35\,{\rm s^{-1}}$, was so small that we could not check whether the power law held. Chen et al. investigated the parameter dependence of the frequency by varying the number of candles, which could correspond to $d$ in the viewpoint of the fuel supply; the frequency was decreased with an increase in the number of bundled candles\cite{Chen2019}. The results in our simulation and experiment were qualitatively consistent with their results.

We experimentally observed the two types of oscillations with different amplitudes and frequencies: oscillatory states A and B. Previously, Cetegen et al. and Bunkwang et al. reported bistability of the axisymmetric and asymmetric states\cite{Bunkwang2020,Cetegen2000}. Although the frequencies in oscillatory states A and B were quantitatively consistent with those in the reported axisymmetric and asymmetric states, respectively, the difference in the flame-shape symmetry between oscillatory states A and B was not observed in our experiment. Therefore, whether oscillatory state B corresponded to the asymmetric state is still unclear, and further investigation is needed in both three-dimensional hydrodynamic simulation and experiments by focusing on the axial symmetry breaking.

\begin{figure*}[tb]
  \includegraphics{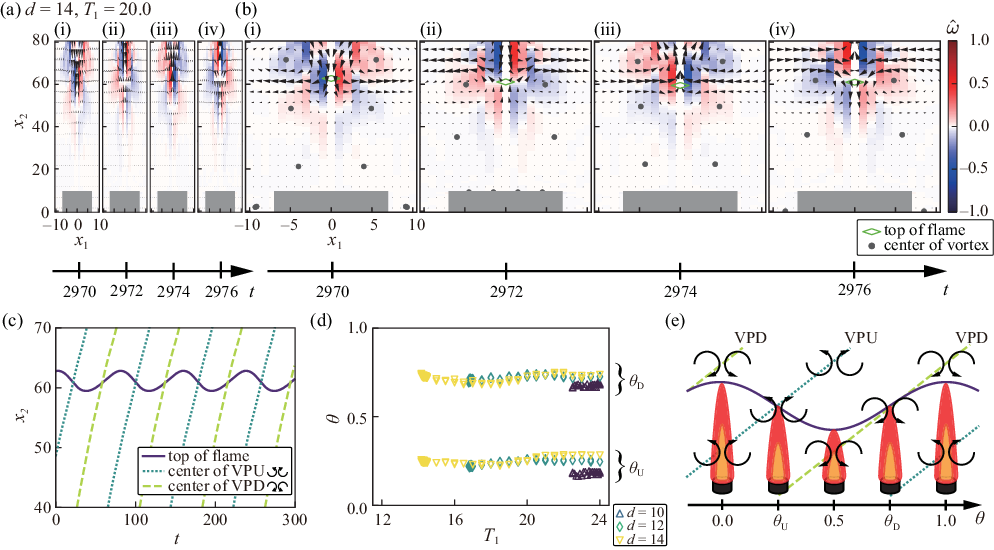}
  \caption{
    (a) Snapshots of the velocity field $\hat{v}_\alpha$, which is defined in Eq.~(\ref{quad:6}), after a sufficiently long time for $d= 14$ and $T_1= 20$. The vorticity field $\hat{\omega}$ is also plotted with the color code. (b) Velocity field and vorticity field enlarged horizontally. The region corresponding to the wick is indicated by gray. The top position of the flame is indicated by a green rhombus. The center positions of the vortex pair are indicated by gray filled circles. (c) Time series of the top position of the flame and the center of the vortex pair. (d) $T_1$- and $d$-dependence of $\theta_{\rm U}$ and $\theta_{\rm D}$, the phases at which VPU and VPD rose, respectively. (e) Schematic illustration of the flame height and vortex structure during one period of the flame oscillation.
  \label{fig:11}}
\end{figure*}

Following some previous studies\cite{Fujisawa2018,Carpio2012,Durox1997,Zhou2001,Yang2019,Bunkwang2020,Tokami2021}, we checked the flow structure in our simulation to understand the role of the large-scale hydrodynamic structure on the flame oscillation. In order to focus on the time variation of the flow in the oscillatory state, we obtained the velocity field $\hat{v}_\alpha$, which is defined as the deviation from the time-averaged flow field as
\begin{align}
  \hat{v}_\alpha=v_\alpha-\frac{1}{\tau}\int_t^{t+\tau}v_\alpha dt,\label{quad:6}
\end{align}
where $\tau$ is the period of the oscillation. The sequential snapshots of $\hat{v}_\alpha$ after a sufficiently long time are shown in Fig.~\ref{fig:11}(a) and (b). The vorticity field,
\begin{align}
  \hat{\omega}=\partial_1\hat{v}_2-\partial_2\hat{v}_1,\label{quad:7}
\end{align}
is also plotted with the color map. We observed the vortex pair with the upward velocity along the center axis (VPU) and that with the downward velocity along the center axis (VPD) rising alternately. Such vortex dynamics was observed in a previous experimental study, in which the velocity field in the flame oscillation was measured with particle image velocimetry and ``invalid velocity vector analysis'' based on the proper orthogonal decomposition\cite{Fujisawa2018}. Note that the vortex pair in the two-dimensional systems corresponds to the vortex ring in the three-dimensional systems.

In order to clarify the effect of the vortex pairs on the oscillation of the flame height, we investigated the timing at which the vortex pair rose. The top position of the flame and the center positions of the vortices, at which $\hat{v}_1=\hat{v}_2=0$ holds, are indicated in Fig.~\ref{fig:11}(b), and the time series of the $x_2$-components are shown in Fig.~\ref{fig:11}(c). The center positions of VPD and VPU passed through the flame top when the flame height went up and down, respectively. We quantified this instant by the phase $\theta$ of the oscillation of the flame height. Here, one period corresponds to $0\leq\theta<1$, and the instant when the flame top is the highest corresponds to $\theta=0$. The phases when VPU and VPD pass through the mean of the flame height are set as $\theta=\theta_{\rm U}$ and $\theta=\theta_{\rm D}$, respectively. The dependences of $\theta_{\rm U}$ and $\theta_{\rm D}$ on $d$ and $T_1$ are shown in Fig.~\ref{fig:11}(d). We found that $\theta_{\rm U}\simeq1/4$ and $\theta_{\rm D}\simeq3/4$ irrespective of the parameters $d$ and $T_1$. Here, we discuss the mechanism of the flame oscillation by focusing on the horizontal flow. An illustration of the behavior during one period of the flame oscillation is shown in Fig.~\ref{fig:11}(e). The horizontal inflow toward the flame decreased the flame height at $\theta=\theta_{\rm U}$, whereas the horizontal outflow increased the flame height at $\theta=\theta_{\rm D}$. Thus, we considered that the horizontal outflow and inflow may supply the hotter and colder gases, which may expand and shrink the flame, respectively, under the flame sheet model. It is notable that the vertical flow did not explain the phase relationship, since the upward and downward flow may supply more and less fuel, respectively. The above discussion supports that VPU and VPD cause the decrease and increase in the flame height, respectively, and may enhance the instability of the stationary flame, which leads to the oscillation. The oscillation of flame height and the vortex pair periodically rising may enhance each other with time delay, which can lead to the hysteresis and bistability.

\section{conclusion}
We investigated the bifurcation structure of the flame oscillation in a 2D hydrodynamical simulation and experiment. In our simulation, we reproduced the flame oscillation and observed the hysteresis and bistability between the stationary and oscillatory states. In our experiments, we constructed a system in which we could finely vary the wick diameter $d$ as a parameter, and we observed the switching between the stationary and oscillatory states. We concluded that the oscillatory state appeared from the stationary state through the subcritical Andronov-Hopf bifurcation in both the simulation and experiment. In addition, the amplitude was increased and the frequency was decreased as the wick diameter $d$ was increased in both the simulation and experiment.

Following the previous studies, we analyzed the flow structure in our simulation and observed the vortex pairs rising periodically; this provides insights into the mechanisms of the flame oscillation. In addition, we also experimentally observed the quasiperiodic oscillation, which has never been reported before, and another oscillatory state, where the frequency was higher. Understanding of these oscillations may contribute to controlling the flame, and thus, they need to be investigated in detail in the future. Identification of the bifurcation structure of the flame oscillation may also facilitate the investigation of synchronization of coupled flame oscillations using mathematical models.

\section{acknowledgments}
This work was supported by JSPS KAKENHI Grant Nos. JP19H00749 and JP21K13891 (HI), and Nos. JP20H02712, JP21H00996, and JP21H01004 (HK), and also the Cooperative Research Program of ``Network Joint Research Center for Materials and Devices: Dynamic Alliance for Open Innovation Bridging Human, Environment and Materials'' (No. 20214004). This work was also supported by JSPS and MESS under the Japan-Slovenia Research Cooperative Program No. JPJSBP120215001 (HI), and JSPS and PAN under the Japan-Poland Research Cooperative Program No. JPJSBP120204602 (HK).
\bibliography{My_Collection}
\end{document}